\documentclass[twocolumn,aps,superscriptaddress,preprintnumbers]{revtex4-1}

\usepackage{amsmath,amssymb}
\usepackage{graphicx}
\usepackage{dcolumn} 
\usepackage{bm} 
\usepackage{soul}
\usepackage{multirow}
\usepackage{setspace}
\usepackage[mathlines]{lineno}
\usepackage{textcomp}
\usepackage{mathcomp}
\usepackage{spverbatim}
\usepackage{float}
\usepackage{wasysym}
\usepackage{array}
\usepackage{subfigure}
\usepackage{physics}
\usepackage{xcolor}

\usepackage{hyperref}
\hypersetup{
    colorlinks=true,       % false: boxed links; true: colored links
    linkcolor=blue,        % color of internal links
    citecolor=blue,        % color of links to bibliography
    filecolor=blue,        % color of file links
    urlcolor=blue,         % color of external links
    runcolor=blue
}

\usepackage{enumitem,amssymb}
\newlist{todolist}{itemize}{2}
\setlist[todolist]{label=$\square$}
\usepackage{pifont}

\newcommand{\USACHP}{Department of Physics, Universidad de Santiago de Chile, Av. Victor Jara 3493, Santiago, Chile;}
\newcommand{\UNDC}{Department of Chemical and Biomolecular Engineering, University of Notre Dame, Notre Dame, Indiana 46556, United States;}
\newcommand{\BCDC}{Department of Chemistry and Biochemistry, University of Central Arkansas, Conway, Arkansas 72035, United States;}
\newcommand{\USACHC}{Department of Materials Chemistry, Universidad de Santiago de Chile, Santiago, Chile;}
\newcommand{\MIRO}{Millennium Institute for Research in Optics, Concepción, Chile.}

\begin{document}
\renewcommand{\figurename}{Figure}
\renewcommand{\tablename}{Table}

\title{Hybrid Atomistic-Parametric Decoherence Model for Molecular Spin Qubits}

\author{Katy Aruachan}
\affiliation{\USACHP}
\author{Sanoj Raj}
\affiliation{\BCDC}
\author{Yamil J. Colón}
\affiliation{\UNDC}
\author{Daniel Aravena}
\affiliation{\USACHC}
\author{Felipe Herrera}
\email{felipe.herrera.u@usach.cl}
\affiliation{\USACHP}
\affiliation{\MIRO}

\date{\today}

\begin{abstract}
Solid-state molecular qubits with open-shell ground states have great potential for addressability, scalability, and tunability, but understanding the fundamental limits of quantum coherence in these systems is challenging due to the complexity of the qubit environment. To address this, we develop a random Hamiltonian approach where the molecular $g$-tensor fluctuates due to classical lattice motion obtained from molecular dynamics simulations at constant temperature. Atomistic $g$-tensor fluctuations are used to construct Redfield quantum master equations that predict the relaxation $T_1$ and dephasing $T_2$ times of copper porphyrin qubits in a crystalline framework. Assuming one-phonon spin-lattice interaction processes, $1/T$ temperature scaling and $1/B^3$ magnetic field scaling of $T_1$ are established using atomistic bath correlation functions. Atomistic $T_1$ predictions overestimate the available experimental data by orders of magnitude. Quantitative agreement with measurements at all magnetic fields is restored by introducing a magnetic field noise model to describe lattice nuclear spins, with field-dependent noise amplitude in the range $\delta B\sim 10\,\mu{\rm T}- 1\,{\rm mT}$ for the copper porphyrin system. We show that while $T_1$ scales as $1/B$ experimentally due to a combination of spin-lattice and magnetic noise contributions, $T_2$ scales strictly as $ 1/B^2$ due to low-frequency dephasing processes associated with magnetic field noise. Our work demonstrates the potential of dynamical methods for modeling the open quantum system dynamics of molecular spin qubits.
\end{abstract}
\maketitle

\section{\label{sec:intro}Introduction} 

Molecular complexes with low-spin electronic ground states are promising platforms for solid-state quantum information processing \cite{Aromi2012,gaita2019molecular,Atzori:2019}. Similar to other platforms, such as semiconductor defects \cite{Stanwix2010,Christle2015}, molecular spin qubits can be optically addressed \cite{Bayliss2020} and coherently coupled to low-dimensional resonators for quantum state manipulation \cite{Mergenthaler2017,Gimeno2023}. The spin coherence time of molecular qubits can be longer than other platforms \cite{zadrozny2015} and further extended using dynamical decoupling \cite{Dai:2021} or by exploiting clock transitions \cite{Shiddiq2016}. A key advantage of molecular systems is the possibility of engineering their environment with angstrom-scale precision using a range of chemical design tools \cite{Graham2017,yu2019concentrated,yu2020spin,amdur2022chemical}. However, future improvements in coherence lifetime of molecular spin qubits require better theoretical and experimental understanding of qubit-reservoir interactions than what is currently available \cite{Lavroff:2021ww}

Multi-scale {\it ab-initio} modeling approaches have been developed and used for computing $T_1$ (relaxation) and $T_2$ (decoherence) times of molecular spin qubits \cite{tesi2016giant,albino2019first,Lunghi2019, lunghi2022toward, lunghi2017role, lunghi2017intra,bayliss2022enhancing,Goh2022,SauzadelaVega2025}, providing atomistic insights on the coupling of the qubit spin with vibrational modes and nuclear spins in the crystal lattice. These insights can potentially be used for developing chemical design rules that optimize qubit performance. An atomistic description based on the electronic structure of spin qubit materials has in principle higher predictive power than phenomenological theories developed for electron paramagnetic resonance (EPR) spectroscopy \cite{VanVleck1940,Roessler2018},  but requires accurate evaluation of a large number of coupling constants that describe the interaction of qubit spins with vibrational normal modes of the lattice or nearby atomic nuclear spins.  Spin-lattice coupling constants are obtained through finite-difference evaluation of high-order derivatives of the electronic Hamiltonian with respect to atomic displacements along the $3N-6$ normal modes of the lattice structure \cite{Lunghi2019,shushkov2024novel,Dmitriev2025}, with atom numbers $N\sim 10^2 $, depending on the unit cell size. In addition to the high computational cost, which could be mitigated with machine learning interpolation \cite{lunghi2022computational}, numerical gradients with respect to normal mode coordinates can introduce errors if the  symmetries of the Hamiltonian are not preserved in the finite difference representation \cite{Li2009}. 

In this work, we develop a hybrid atomistic-parametric methodology for constructing Redfield quantum master equations that predict $T_1$ and $T_2$ times of molecular spin qubits embedded in solid-state crystal frameworks. The method builds on a previous stochastic Hamiltonian formulation of the open quantum system problem \cite{aruachan2023semi}. In this formulation, the influence of the environment on the qubit Zeeman Hamiltonian is partitioned into stochastic fluctuations of the gyromagnetic tensor $\delta g_{ij}(t)$, due to lattice phonons and molecular vibrations, plus stochastic fluctuations of the local magnetic field at the qubit location $\delta B_i(t)$, due to induced fields produced by nuclear spins distributed throughout the lattice \cite{Schwartz1955,Jensen1973} or vibrational orbital angular momentum \cite{shushkov2024novel}. To construct the  bath spectral density $J(\omega)$, which determines the qubit relaxation and decoherence timescales, we obtain $\delta g_{ij}(t)$ from first principles by sampling the electronic effective spin Hamiltonian of the qubit over an ensemble of molecular dynamics trajectories at constant temperature. No numerical derivatives of the Hamiltonian are evaluated. Spectral densities from dynamical quantum-classical dynamics have been used for optical spectroscopy \cite{Valleau2012}. Here we adapt the approach for solid-state electron spin dynamics for the first time.

The rest of the paper is organized as follows: Section \ref{sec:model} describes the open quantum system model and the semi-classical approach for constructing the total spectral density. Section \ref{sec:results} shows results for $T_1$ and $T_2$ as functions of magnetic field for copper porphyrin qubits embedded in a crystal framework. Section \ref{sec:conclusions} presents conclusions and perspectives for model  improvements and applications.

\section{Open Quantum System Model}
\label{sec:model}

We model  the relaxation dynamics of a single molecular spin qubit using Haken–Strobl theory \cite{haken1972coupled,rips1993stochastic}. The system is composed of an unpaired electronic spin ($S=1/2$) in an external magnetic field $\mathbf{B}=\sum_jB_j\mathbf{e}_j$. The molecular spin is embedded in a solid-state matrix and couples to lattice phonons and other nuclear spins present in the crystal structure. In Haken-Strobl theory, bath degrees of freedom are treated implicitly via time-dependent fluctuations that they induce on the total system Hamiltonian $\hat{H}(t) = \hat{H} _{S} +\hat{H}_{SB}(t)$, where $\hat{H}_{S}$ is time-averaged over bath-fluctuation timescales. $\hat{H}_{SB}(t)$ contains the explicit fluctuation variables. For $S=1/2$ molecules, $\hat{H}_{S}$ is given by \cite{stoll2006easyspin}
\begin{equation}\label{eq:system-H}
\hat{H}_{S}  = \frac{1}{2}\mu_{B}\, g_{ij} B_{i}\hat \sigma_{j} + \frac{\hbar}{2}A_{ij} \,\hat \sigma_{i}\hat I_{j},          
\end{equation}
where $B_{i}$ are external magnetic field components, $g_{ij}$ are elements of the gyromagnetic $g$-tensor,  $\hat{\sigma}_{j}$ are the Pauli matrices, $\hat{I}_{j}$ are nuclear spin angular momentum components of the qubit molecule, and $A_{ij}$ are elements of hyperfine tensor. 

Bath-induced fluctuations have the form \cite{aruachan2023semi}
\begin{equation}\label{eq:system-bath-H}
\hat{H}_{SB}(t)  = \frac{1}{2}\mu_{B}\delta g_{ij}(t) B_{i}\hat \sigma_{j} +\frac{1}{2}\mu_{B} \delta B_{i}(t) g_{ij}\hat \sigma_{j},     
\end{equation}
where $\delta g_{ij}(t)$ arises from lattice motion and vibrations that modifies the qubit response to external magnetic fields and $\delta B_{i}(t)$ corresponds to random  local magnetic fields produced by nearby nuclear spins in the crystal lattice \cite{Freitas2015} or effective magnetic fields due to vibrational orbital angular momenta \cite{shushkov2024novel}. Fluctuations of the hyperfine tensor $A_{ij}$ are neglected to lowest order and $\langle \delta g_{ij}\,\delta B_k\rangle =0$ at all times.

\begin{figure*}[t]
    \includegraphics[width=\textwidth]{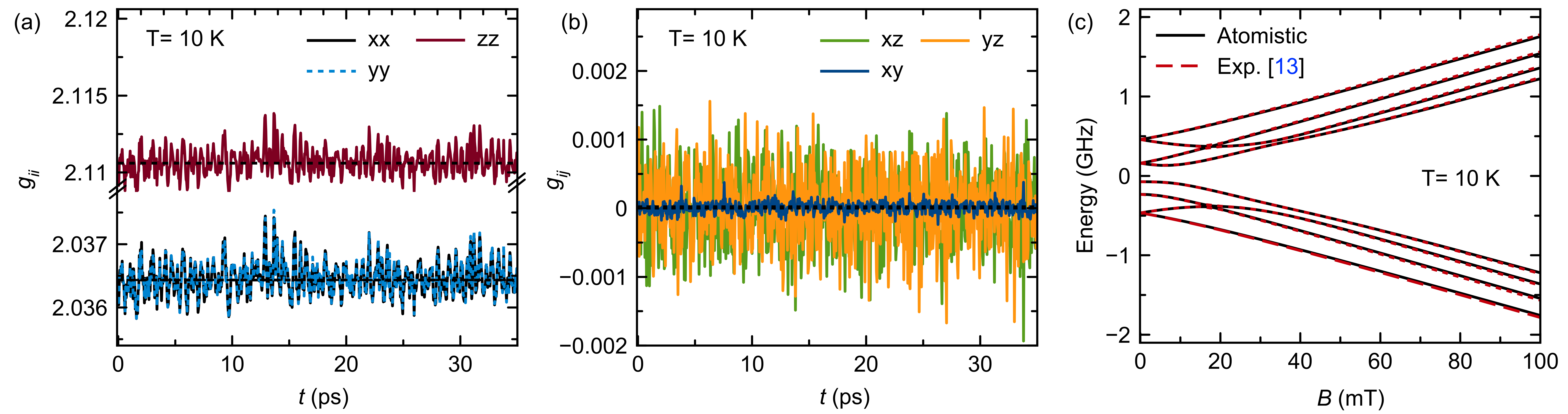}
    \caption{{\bf Atomistic $g$-tensor fluctuations.} (a) Evolution of the diagonal tensor elements $g_{ii}(t)$ of copper qubits ($S=1/2, I=3/2$) in metal-organic frameworks (Cu-PCN-224) at 10 K, sampled from equilibrium molecular dynamics simulation. Time-averaged values are shown in dashed horizontal lines. (b) Off-diagonal elements $g_{ij}$ around their zero-valued time averages at 10 K. (c) Zeeman spectrum of the copper qubit obtained from atomistic time-averaged $g$-tensor elements at 10K, compared with experimental data from Ref. \cite{yu2019concentrated}.}
  \label{fig:g tensor fluctuations}
\end{figure*}

The qubit coherence and dephasing times, $T_1$ and $T_2$, are obtained from the evolution of system density matrix $\hat{\rho}_{S}$. In the interaction picture with respect to $\hat{H}_{S}$, the matrix elements $\rho_{ab}=\langle a| \hat{\rho}_{S} | b\rangle $ in the basis of system eigenstates $\ket{a}$ evolve according to the Redfield quantum master equation \cite{breuer2002theory}
\begin{equation} \label{eq:Redfield equation}
\frac{d}{dt}\rho_{ab}(t) = -\sum_{c,d}R_{ab,cd}\rho_{cd}(t).
\end{equation} 
where $R_{ab,cd}$ are elements of the Redfield tensor encoding the relaxation and decoherence dynamics of the spin qubit. In terms of state-to-state transition rates, Redfield tensor elements can be written as \cite{may2023charge}
\begin{eqnarray}\label{eq:Redfield tensor}
R_{ab,cd} &= & \frac{1}{\hbar^{2}}\left\{\delta_{bd}\sum_{e}\Gamma_{ae,ec}(\omega_{ce}) -  \right. \Gamma_{ca,bd}(\omega_{db}) \nonumber\\ 
&  & \left.- \Gamma_{db,ac}(\omega_{ca}) + \delta_{ac}\sum_{e}\Gamma_{be,ed}(\omega_{de}) \right\} 
\end{eqnarray}
where $\Gamma_{ab,cd}(\omega_{dc})$ are decay rate functions evaluated at the system transition frequencies $\omega_{ab} = \omega_{a} - \omega_{b}$. The rates are in turn given by
\begin{equation}\label{eq:Gamma}
\Gamma_{ab,cd}(\omega_{dc}) = {\rm Re}\left[\langle a |\hat{\sigma}_{j}|b\rangle \langle c |\hat{\sigma}_{j'}| d \rangle J_{jj'}(\omega_{dc})\right]
\end{equation} 
where $J_{jj'}(\omega)$ (units of ${\rm J}^2\,{\rm Hz}^{-1}$) is the bath spectral density representing the strength of the coupling to the environment at a given system transition frequency and temperature. Following Ref. \cite{aruachan2023semi}, the total spectral density $J_{jj'}(\omega)$ is partitioned as 
\begin{equation}\label{Eq:specTol}
J_{jj'}(\omega) = J_{jj'}^{\delta g}(\omega)  + J_{jj'}^{\delta B}(\omega), 
\end{equation}
where $J_{jj'}^{\delta g}(\omega)$ is the spectral density associated with $g$-tensor fluctuations and $J_{jj'}^{\delta B}(\omega)$ the spectral density due to  local magnetic noise. 

To obtain $T_{1}$ and $T_{2}$ at a given magnetic field strength and temperature, the spectral densities  $J_{jj'}^{\delta g}$ and $J_{jj'}^{\delta B}$ are obtained as described below, and Redfield tensor elements are constructed from Eqs. (\ref{eq:Redfield tensor}) and (\ref{eq:Gamma}) to solve Eq. (\ref{eq:Redfield equation}) for a pure non-equilibrium initial state  $\hat{\rho}_{S}(0)=\ket{\psi}\bra{\psi}$. The fully polarized spin-up state $\ket{\psi}=\ket{m_s=+1/2}\ket{m_I}$ is used to obtain $T_1$ from the decay dynamics of $\langle \hat{S}_{z}(t)\rangle$, and the $x$-polarized state $\ket{\psi}=(1/\sqrt{2})(\ket{m_s=+1/2}+\ket{m_s=-1/2})\otimes \ket{m_I}$ is used to obtain $T_2$ from the decay dynamics of $\langle \hat{S}_{x}(t)\rangle$. Other initial conditions lead to small changes in $T_1$ and $T_2$ that do not affect the order magnitude or scaling with field strength and temperature. Different initial nuclear spin states $\ket{m_I}$ also give similar decay timescales. 

Fluctuations of the $g$-tensor due to phonons can be obtained from classical dynamics simulations of the lattice dynamics and classical autocorrelation functions can be defined as (units of ${\rm Hz}^{-1}$)
\begin{equation}\label{eq:Gdg}
G^{\delta g}_{ij}(\omega) = \frac{1}{\sqrt{2\pi}} \int_{0}^{\infty} dt \langle \delta g_{ij}(t) \delta g_{ij}(0)\rangle e^{\mathrm{i}\omega t}.
\end{equation}
Classical bath autocorrelation functions are in general temperature dependent, but do necessarily lead to spectral densities that satisfy  detailed balance conditions $\Gamma(\omega)/\Gamma(-\omega)={\rm exp}[\hbar\omega/k_{\rm B}T]$ for Eq. (\ref{eq:Gamma}). Schemes for constructing quantum spectral densities from classical autocorrelation functions are known  \cite{egorov1999quantum,Valleau2012}. We adopt a harmonic approximation to define the spin-lattice spectral density as
\begin{equation}\label{eq:dg-spectral}
J_{jj'}^{\delta g}(\omega) = \delta_{jj'}\left(\mu_{B}/2\right)^{2}\,(\lambda \omega \,B_{i}^2) \, G^{\delta g}_{ij}(\omega),
\end{equation}
where $\lambda$ is an Ohmic parameter set throughout this work as $1/\lambda = 6.9\,{\rm cm}^{-1}$. 

Classical autocorrelation functions (ACF) $C_{ij}(t,t')\equiv\delta g_{ij}(t)\delta g_{i j}(t')$ for particular qubit systems can be defined by computing $g$-tensors for an ensemble of crystal snapshots sampled from equilibrium molecular dynamics (MD) simulations. In this work, we focus on modeling the copper porphyrin qubit system Cu-PCN-224, where the qubit spin is embedded in a metal-organic framework (MOF) crystal. Copper porphyrin transition metal ions Cu$^{+2}$ ($S=1/2$, $I = 3/2$) are well characterized  \cite{haakansson2013cu} and in Cu-PCN-224 the copper ion is dilute enough in the lattice to prevent electron spin-spin interactions.  EPR spectroscopy and relaxation measurements are available for this qubit system at different magnetic fields and temperatures \cite{yu2019concentrated}.  

 MD simulations were performed with Large-scale Atomic/Molecular Massively Parallel Simulator (LAMMPS) \cite{thompson2022lammps} in the NVT ensemble with a time step of $1$ fs. The Lammps-interface \cite{boyd2017force} was used to prepare the input data files using UFF4MOF \cite{addicoat2014extension, rappe1992uff} force field parameters. A non-bonded cutoff of 12.5 Å was applied, with Lennard-Jones potential for the non-bonded interactions and cross-terms were modeled using Lorentz-Berthelot mixing rules. Five simulation replicas were performed at $10$ K, $50$ K, $100$ K, $150$ K and $300$ K, each simulation running for $1$ ns with frames extracted every $50$ fs. Crystal geometry snapshots taken every 25 steps from the MD simulations were used to compute instantaneous $g$-tensors using density functional theory (DFT) with the B3LYP functional and Def2-TZVP basis set. Interpolation was used to join results from different snapshots over a total simulation time of approximately $900$ ps.

\begin{figure}[t]
   \centering   
   \includegraphics[width=0.45\textwidth]{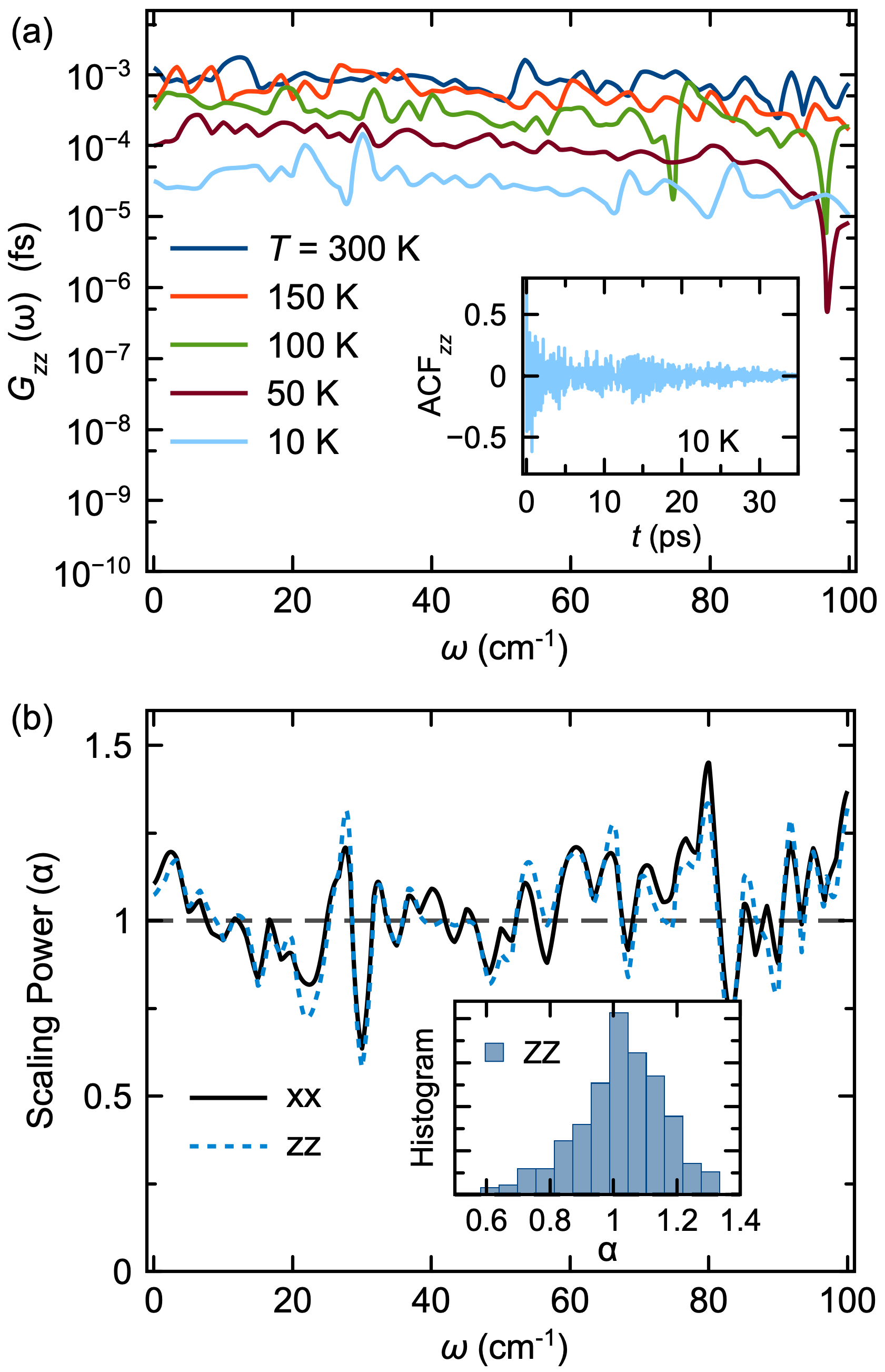}
   \caption{{\bf Spectrum of $g$-tensor fluctuations.} (a) $G_{zz}$ fluctuation spectrum at different temperatures. The inset shows a sample autocorrelation function (ACF) for $zz$ fluctuations at 10 K. (b) Temperature scaling power $\alpha$ in $G_{ii}\sim T^\alpha$ for $xx$ and $zz$ components, evaluated at different frequencies. The inset shows the histogram of $\alpha$ values for the $zz$ component} \label{fig:G spectrum}
\end{figure} 

Local magnetic field noise due to nuclear spins in the lattice is responsible for pure dephasing and low-frequency relaxation of the qubit ($\omega\rightarrow 0$) and therefore is not subject to detailed balance conditions. The spin noise spectral density can be written as
\begin{equation}\label{eq:dB-spectral}
J_{jj'}^{\delta B}(\omega) = \left(\frac{\mu_{B}}{2}\right)^{2}g_{ij}g_{i'j'} G^{\delta B}_{ii'}(\omega),
\end{equation}
with magnetic noise spectrum given by (units of ${\rm T}^2\,{\rm Hz}^{-1}$)
\begin{equation}\label{eq:GdB}
G^{\delta B}_{ii'}(\omega) = \frac{1}{\sqrt{2\pi}}\int_{0}^{\infty} dt \langle \delta B_{i}(t) \delta B_{i'}(0)\rangle e^{\mathrm{i}\omega t}.
\end{equation}
We model the autocorrelation function of the random spin noise as being invariant under rotations (isotropic), of the form  
\begin{equation}\label{eq:AutoB model}
\langle\delta B_{i}(\tau)\delta B_{i'}(0)\rangle  \equiv A_{B}(B)\,e^{-\gamma_{\rm pd} \tau} \delta_{ii'},
\end{equation}
with dephasing rate $\gamma_{pd}$ and field-dependent noise strength $A_{B}$ given by \cite{aruachan2023semi}
\begin{equation}\label{eq:AofB model}
A_{B}(B) = a +b\,B^{2},
\end{equation} 
where $a$ and $b$ are constants and $B$ is the external magnetic field strength. The magnitude of local magnetic noise is $\delta B \equiv \sqrt{A_{B}}$. This model is  consistent with classical linear response  $\delta B\sim \sqrt{b}B$ at moderate fields \cite{kittel2005introduction}. The model is parametrized by comparing with available qubit relaxation data.

\section{Results and Discussion}
\label{sec:results}

%\subsection{Spectrum of $g$-tensor fluctuations}

Figure \ref{fig:g tensor fluctuations}a and \ref{fig:g tensor fluctuations}b show the evolution of the diagonal and off-diagonal $g$-tensor elements $g_{ij}(t)$ of  Cu-PCN-224 at $T=10$ K, computed as described above. By construction, the fluctuations preserve the anisotropy of the $g$-tensor and the symmetry of the Hamiltonian. By time-averaging the curves over the simulation interval (900 ps), we obtain the stationary elements $\langle g_{xx}\rangle_t= \langle g_{yy}\rangle_t=g_{\perp} = 2.1106$, $\langle g_{zz}\rangle_t =g_{\parallel} = 2.0364$ and $\langle g_{xy}\rangle_t=\langle g_{xz}\rangle_t=\langle g_{yz}\rangle_t\ll10^{-5}$, where $\langle X\rangle_t$ denotes time average. The time-averaged $g$-tensor is used to construct the Zeeman term in Eq. (\ref{eq:system-H}). Hyperfine parameters are not computed from atomistic simulations, but extracted from EPR measurements in Ref. \cite{yu2019concentrated} as $A_{zz} = 611$ MHz and $A_{xx} = A_{yy} = 79.4$ MHz. Figure \ref{fig:g tensor fluctuations}c shows the energies of the Zeeman sub-levels obtained by combining the time-averaged atomistic $g$-tensor with empirical hyperfine parameters. The spectrum agrees with spectroscopic measurements up to minor differences in transition frequencies  on the order of $10^{-2}$ cm$^{-1}$.

The fluctuations $\delta g_{ij}(t) \equiv g_{ij}(t) - \langle g_{ij}\rangle_t$ are used to construct an ensemble of classical autocorrelation functions $C_{ij}(t,t')$ by segmenting the simulation interval into non-overlapping windows of 35 ps duration each. By averaging over a large number of autocorrelation windows, stationary classical autocorrelations $\langle \delta g_{ij}(\tau)\delta g_{ij}(0) \rangle$ are obtained. Figure \ref{fig:G spectrum}a shows the simulated spectrum of $zz$ fluctuations, $G_{zz}^{\delta g}(\omega)$, at different simulation temperatures. The inset shows a representative autocorrelation function $C_{zz}(t,t')$ over an integration window. The noise level overall increases monotonically with temperature, and frequency noise of the fluctuation data is significant despite the large number of autocorrelation windows used. Similar results are found for other components of the fluctuation tensor $G_{ij}^{\delta g}$. 

To understand the temperature scaling $G_{ij}\sim T^\alpha$, we analyze the spectrum of $G_{xx}^{\delta g}$ and $G_{zz}^{\delta g}$ at different temperatures to obtain the pseudo-spectrum of $\alpha$, shown in Fig. \ref{fig:G spectrum}b. Up to $\omega\approx 100\, {\rm cm}^{-1}$, we obtain the scaling factors $\alpha= 1.03 \pm 0.13$ for $G_{xx}$ and $\alpha = 1.02 \pm 0.14$ for $G_{zz}$. Other spectral components $G_{ij}$ also give $\alpha _{ij}\approx 1.0$. Since the Redfield relaxation rates are directly proportional to $G_{ij}$ [Eq. (\ref{eq:Gamma})], linear scaling gives the temperature dependence $T_1\sim 1/T$, which is characteristic of direct absorption and emission of phonons at the qubit transition frequency \cite{VanVleck1940,Sauter2021}. 

%\subsection*{Relaxation and dephasing times for copper porphyrin qubits}

  \begin{figure}[t]
    \centering
   \includegraphics[width=0.45\textwidth]{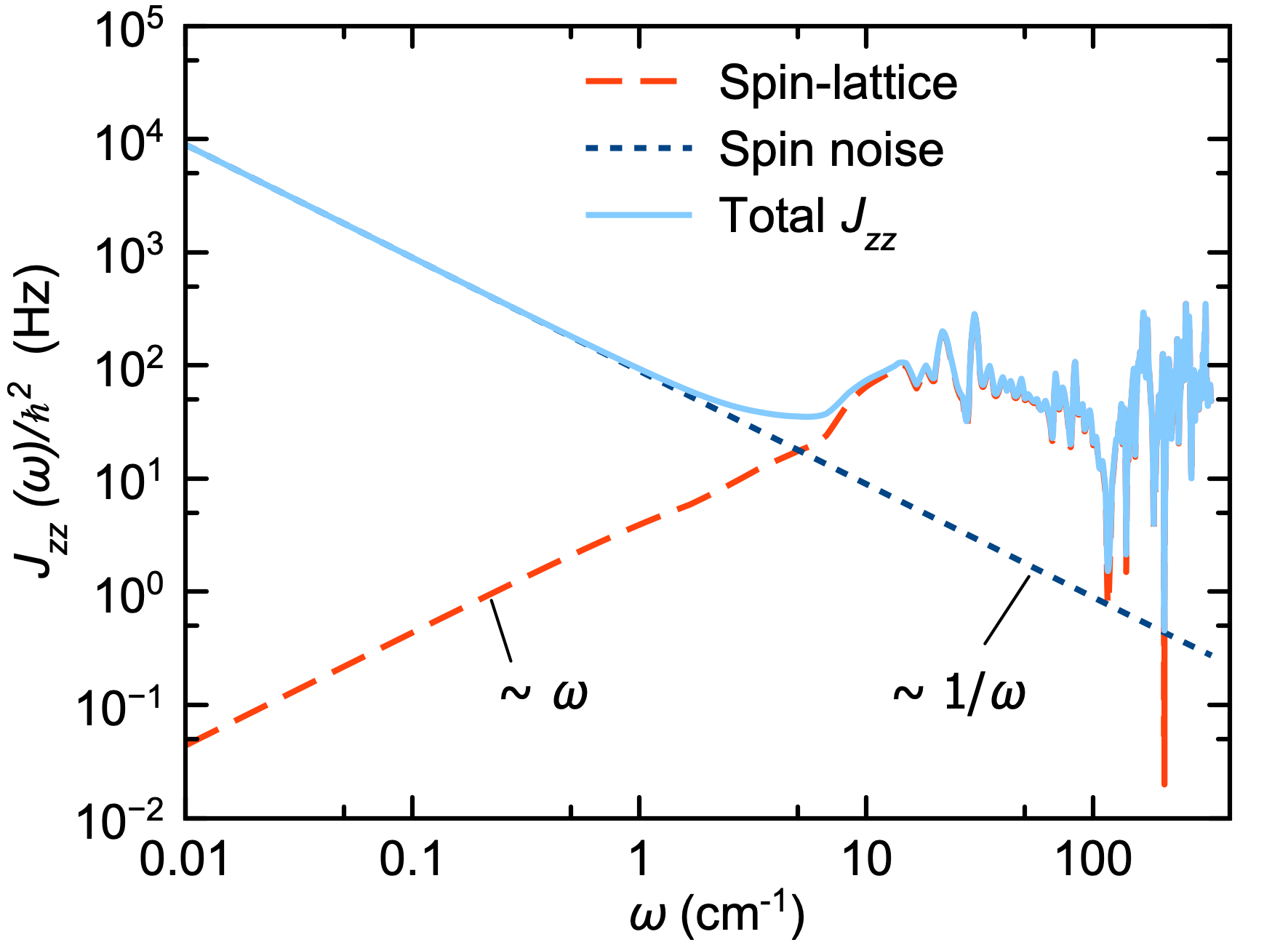}
    \caption{{\bf Hybrid spectral density for copper qubits.} Total spectral density $J_{zz}$ (solid line) obtained by combining the atomistic spectral density $J_{zz}^{\delta g}$ at $10$ K (dashed line) with a model magnetic noise spectral density $J^{\delta B}_{zz}$ having field-independent noise amplitude (dotted line; $\gamma_{\rm pd} = 0.001$ cm$^{-1}$, $a = 16\times 10^{-10}$ T$^{2}$, $b = 0$). }
    \label{fig:spectral densities}
\end{figure}

Figure \ref{fig:spectral densities} shows the total spectral density $J_{zz}$ used to describe the coupling of Cu-PCN-224 qubits with the lattice and spin reservoir at $T=10$ K. The total spectral density is obtained by combining the atomistic spin-lattice spectral density $J_{ij}^{\delta g}$ with fluctuation spectra in Fig. \ref{fig:G spectrum}, with a model magnetic field noise spectral density $J^{\delta B}$ given by Eq. (\ref{eq:dB-spectral}) with parameters set as $\gamma_{\rm pd} = 0.001$ cm$^{-1}$, $a = 16\times 10^{-10}$ T$^{2}$ and $b = 0t$. At high frequencies $\omega\gg \gamma_{\rm pd} $, magnetic field noise is suppressed with increasing frequency as $\sim1/\omega$ due to its Lorentzian character. At low frequencies $\omega\ll \lambda$, coupling to lattice phonons is suppressed due to Ohmic behavior. At intermediate frequencies there is a system-dependent crossover region in which both types of baths contribute roughly equally to the relaxation dynamics \cite{aruachan2023semi}. In our hybrid atomistic-parametric model, at any given temperature the crossover frequency region is controlled by the field noise amplitude parameter $a$.

\begin{figure}[t]
  \includegraphics[width=0.45\textwidth]{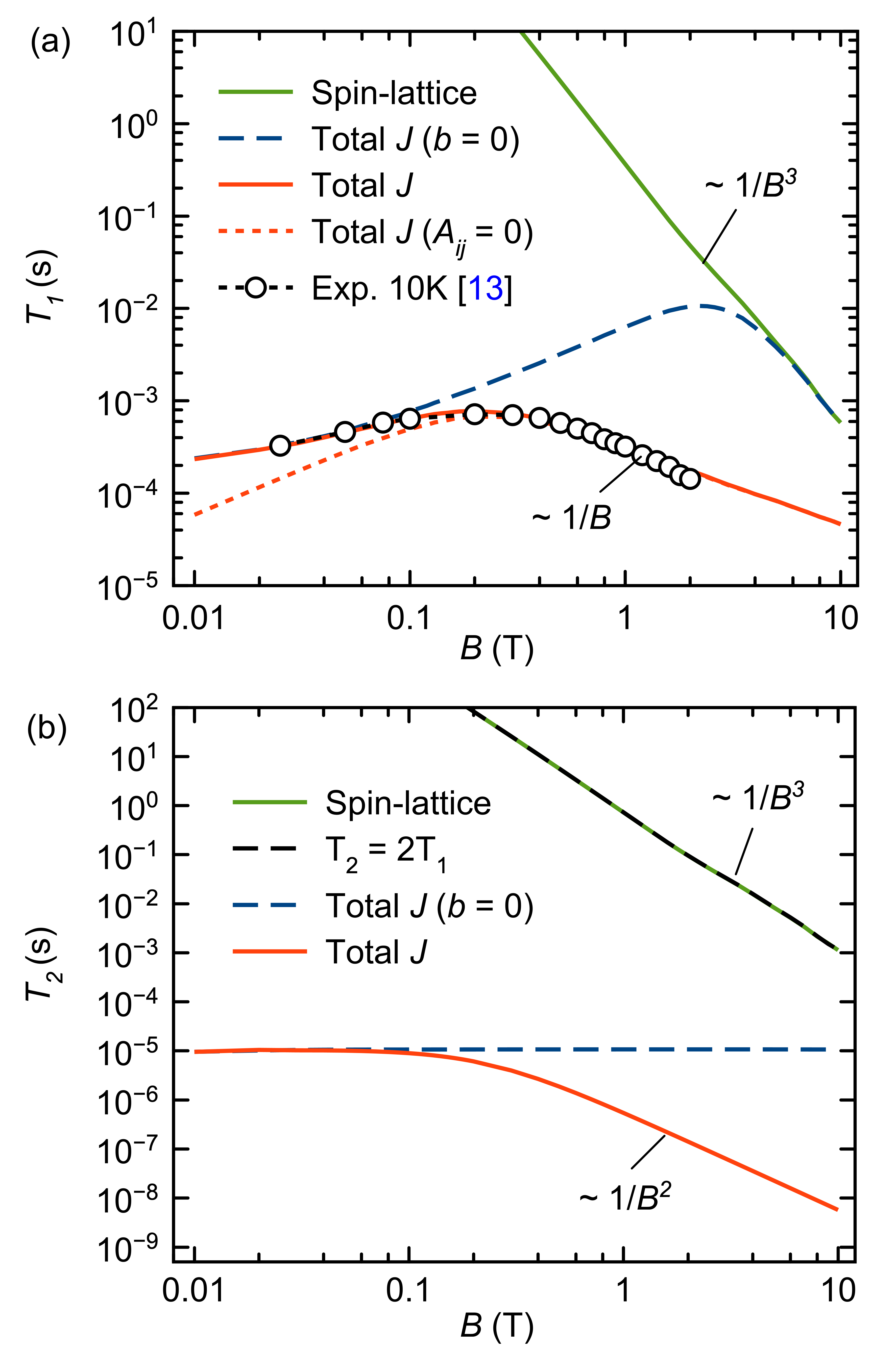}
   \caption{{\bf Relaxation and dephasing of copper qubits.} (a) $T_{1}$ as a function of field strength $B$ for different spectral density models: atomistic spin-lattice  at 10 K (green solid) and hybrid-atomistic  with field-independent (blue dashed) and field-dependent magnetic noise (red solid). Results for a modified Hamiltonian without hyperfine interactions ($A_{ij}=0)$ are also given. Experimental results from Ref. \cite{yu2019concentrated} at 5 K are shown for comparison. (b) $T_2$ as a function of field strength for the models in panel (a). The pure relaxation limit $T_2=2T_1$ is shown for comparison (black dashed line). $\gamma_{\rm pd} = 0.001$ cm$^{-1}$  for all magnetic noise models.} 
  \label{fig:T1 times}
\end{figure} 

Figure \ref{fig:T1 times}a shows $T_1$ predictions for Cu-PCN-224 qubits as a function of magnetic field strength, for different spectral density models. For comparison,  experimental measurements from Ref. \cite{yu2019concentrated} at 5 K are reproduced ($T = 10$ K data was not available). Results show that taking only the atomistic spin-lattice spectral density $J^{\delta g}$ based on the $g$-tensor fluctuations in Fig. \ref{fig:G spectrum}a (10 K) overestimates $T_1$ by at least an order of magnitude at the highest magnetic fields ($B\sim 10 \,{\rm T}$). Since $\omega\sim B$ and $J^{\delta g}\sim  \omega B^2$ [Eq. (\ref{eq:dg-spectral})], the atomistic spin-lattice model predicts a $\sim 1/B^3$ divergence of $T_1$ at vanishing magnetic fields. This zero-field divergence is removed by adding the contribution of magnetic field noise $J^{\delta B}$, which in our model depends on the noise amplitude parameters $(a, b)$ [Eq. (\ref{eq:AofB model})].  By assuming a field-independent magnetic noise amplitude ($b=0$), a crossover field around $B\sim 2.6$ T is established at which $T_1$ is maximum (Fig. \ref{fig:T1 times}a, blue dashed line). The measured $T_1$ values at low fields are reproduced with noise amplitude $\delta B\equiv \sqrt{a}= 40\, \mu {\rm T}$. The magnetic field that gives maximum $T_1\approx 670\,\mu{\rm s}$ in experiments is relatively low ($\approx 0.2$ T), far below the crossover field predicted by the $b=0$ noise model. By introducing field-dependence of the noise amplitude with $b= 3\times 10^{-8}$, the model predictions for $T_1$ follow closely the experimental measurements across the entire range of magnetic fields (Fig. \ref{fig:T1 times}a, red solid line). In this case the field noise amplitude varies from $\delta B\approx40\,\mu{\rm T}$ at low fields ($10\, {\rm mT}$) to $\delta B\approx1.7\,{\rm mT}$ at high fields ($10\,{\rm T}$). This magnetic noise model also captures well the measured field scaling $T_1\sim 1/B$, suggesting that the role of spin-lattice relaxation for Cu-PCN-224 qubits is minimal at these magnetic fields. We also show that $T_1$ values computed using the same total spectral density but without considering hyperfine splittings in the qubit spectrum (Fig. \ref{fig:T1 times}, red dashed line) can differ significantly from the full model, at the lowest magnetic fields.

Figure \ref{fig:T1 times}b shows the dephasing time $T_{2}$ as a function of field strength for the same bath models in Fig. \ref{fig:T1 times}a. Phase memory times  from low-field Hahn-echo measurements of  Cu-PCN-224 are in the range $T_m\sim 20-50 \,{\rm ns}$  up to 80 K \cite{yu2019concentrated}. These values roughly set the order of magnitude expected for $T_2$. As discussed in previous work \cite{aruachan2023semi}, the spin-lattice spectral density $J^{\delta g}$ gives $T_2$ values that follow closely the pure relaxation limit $T_{2} = 2T_{1}$ (Fig. \ref{fig:T1 times}b, black dashed), because there is no contribution to pure dephasing at $\omega\rightarrow 0$ for Ohmic spectral densities. Magnetic field noise gives dephasing contributions that can reduce $T_2$ relative to $T_1$ by orders of magnitude. As expected, the magnetic noise model with field-dependent noise amplitude (Fig. \ref{fig:T1 times}b, red solid line) gives $T_2$ values that monotonically decrease with field strength up to $T_2\sim 10\,{\rm ns}$ at the highest fields ($\sim 10$ T). The field dependence of this model scales as $T_2\sim 1/B^2$, showing that only near-zero frequency transitions $(\omega\ll\gamma_{\rm pd})$ are coupled by dephasing channels. Field-independent noise amplitude (Fig. \ref{fig:T1 times}b, blue dashed line) gives $T_2$ roughly constant around $9\,\mu{\rm s} $. All models predict values of $T_2$ in the range $\sim 0.001-10$ $\mu$s depending on field strength, which at low fields tend to be higher than the measured $T_m$ times, although specific pulse sequence modeling should be done for direct comparisons.

%
%----------------------------------------------------------------------------------------%
%			Conclusions and outlook                   		                %
%----------------------------------------------------------------------------------------%
\section{Conclusions and outlook}
\label{sec:conclusions}

We introduced hybrid atomistic-parametric model spectral densities to describe the open quantum system dynamics of copper porphyrin qubits ($S=1/2$, $I=3/2$) embedded in a crystalline framework. These spectral densities are used to construct Redfield quantum master equations that predict relaxation and dephasing times ($T_1,T_2$) as functions of magnetic field strength and lattice temperature.  The atomistic component of the spectral density is obtained through $g$-tensor autocorrelation functions computed from molecular dynamics simulations and density functional calculations that simulate the variation of the qubit electronic structure due to lattice phonons within the Born-Oppenheimer approximation. This dynamical approach avoids the computationally expensive evaluation of numerical Hamiltonian derivatives with respect to normal mode coordinates and is therefore expected to find use in the modeling of other molecular spin qubit systems. Atomistic calculations give Zeeman spectra in good agreement with experiments \cite{yu2019concentrated}, but $T_1$ values that overestimate measurements by orders of magnitude at intermediate ($\sim 1-10$ T) and low ($\sim 0.01-0.1$ T) magnetic fields. The gap between atomistic modeling and experiments can be accurately bridged by introducing a simple isotropic nuclear spin bath model that couples low-frequency transitions and captures pure dephasing contributions to decoherence. The magnetic field noise amplitude associated with this spin bath is in the range $\delta B\sim 10-10^3\,\mu{\rm T}$, consistent with experimental measurements for similar molecular qubit systems \cite{Shiddiq2016}. 

The theory developed here considers direct (second-order) low-frequency exchange of phonon excitations at qubit transition frequencies. Fourth-order spin-lattice interaction including Raman and Orbach processes have been shown to dominate the relaxation dynamics for specific molecular systems \cite{Ding2016,lunghi2022toward,shushkov2024novel}, but the relative contribution of higher-order spin-phonon processes should be properly assessed for each qubit system. For the copper porphyrin qubits studied here, Raman-type spin-lattice interaction involving far-detuned vibrational modes of a few hundred cm$^{-1}$ (typical Larmor frequency is $\omega_0\ll 10 \,{\rm cm}^{-1}$) would give relaxation rates scaling with magnetic field as $T_1\sim [J(\omega)]^2\sim 1/B^4$, ignoring the low-frequency Ohmic behavior of the spectral density $J(\omega)$. However, the experimental data in Fig. \ref{fig:T1 times} shows a much softer magnetic field scaling $T_1\sim 1/B$. Further improvements of the predictive power of atomistic simulations must therefore consider the combined influence of higher-order spin-phonon interaction and electron-nuclear spin-spin interactions over a sufficiently large crystal supercell. By avoiding the evaluation of numerical derivatives as suggested in this work, the computational cost of such theoretical analysis could be significantly reduced.

%----------------------------------------------------------------------------------------%
%				   Acknowledgments	             		                %
%----------------------------------------------------------------------------------------%
\begin{acknowledgments}
KA is supported by ANID through Beca Doctorado 21220245. YC thanks the University of Notre Dame for financial support through startup funds. DA thanks ANID Fondecyt Regular 1210325 for support and Powered@NLHPC: this research was partially supported by the supercomputing infrastructure of the NLHPC (ECM-02). FH  is supported by ANID through Fondecyt Regular 1221420, Millennium Science Initiative Program ICN17\_012. Work supported by the Air Force Office of Scientific Research under award number FA9550-22-1-0245. 
\end{acknowledgments}

%----------------------------------------------------------------------------------------%
%				    	References	             		                %
%----------------------------------------------------------------------------------------%
\bibliographystyle{apsrev4-1}
\bibliography{hybrid-model}

\end{document}